\documentclass[showpacs,twocolumn,superscriptaddress,10pt]{revtex4-1} 
\usepackage{amsmath,amssymb,graphicx}
\begin{document}

\title{Mitigation of the spectral dependent polarization angle response for achromatic half-wave plate}
\author{Tomotake Matsumura}
\email{Corresponding author: tmatsumu@astro.isas.jaxa.jp}
\affiliation{Institute of Space and Astronautical Science (ISAS), Japan Aerospace Exploration Agency (JAXA) \\ 3-1-1 Yoshinodai, Chuo, Sagamihara, Kanagawa 252-5210, Japan. }

\begin{abstract}
Polarimetry using a half-wave plate (HWP) modulator provides the strong tools to avoid a detector $1/f$ noise and instrument-originated spurious polarization systematic effects. While the Pancharatnam achromatic HWP (AHWP) is commonly used for an application that needs a broadband frequency coverage, this technique introduces a frequency-dependent polarization angle rotation. In this paper we propose a new technique to mitigate this effect by introducing a second set of an AHWP. One rotational and one stationary set of AHWPs achieve a broadband coverage of modulation efficiency without the frequency-dependent polarization angle rotation. We conducted measurements by using three layers of sapphire wave plates and demonstrated this technique at millimeter wavelengths between 72 and 162~GHz. We also discuss a potential application in the CMB polarization experiment based on numerical simulations. 
\end{abstract}


\maketitle 

\section{Introduction}
Measurements of the cosmic microwave background radiation (CMB) have been playing an important role to establish the $\Lambda$CDM cosmology. While vast amount of information is learned by the Planck satellite using the temperature anisotropy of the CMB~\cite{planck}, there is a community-wide effort to measure the polarization of CMB to test the inflationary paradigm and to probe the evolution of the universe via interactions between the CMB and gravitational potential from the large scale structures~\cite{Kamionkowski}. These paradigms started to be disclosed by the recent results from the CMB polarization experiments, BICEP2, POLARBEAR, and SPTpol~\cite{bicep2,polarbear_BB,sptpol}. 

Forthcoming CMB polarization experiments require a polarimeter that is free from a detector $1/f$ noise and controls the instrument-originated polarization systematic effects. A polarimeter using a half-wave plate (HWP) modulator provides two attractive features, i) avoiding the detector $1/f$ by modulating and demodulating the signal frequency, and ii) eliminating the detector differencing, which is a source of the instrumentally induced spurious polarization effect, to reconstruct the incident polarization state~\cite{hhz}. MAXIPOL was the first CMB experiment that employed the continuously rotating HWP. Currently a number of CMB experiments, including ABS, EBEX, LiteBIRD, POLARBEAR-1, POLARBEAR-2, QUBIC, SPIDER, SWIPE are pursuing this technology~\cite{maxipol,brad_thesis,abs,ebex,litebird,polarbear1,polarbear2,qubic,spider,swipe}.

While the use of HWP becomes a popular polarimetry technique, a polarimetry using a single HWP limits the use of the electromagnetic frequency range, and thus the observing detection bandwidth. The typical available bandwidth with a single HWP is $\delta \nu/\nu \sim 0.3$ in order to maintain the linear-to-linear polarization conversion efficiency to be more than 0.9. Upcoming CMB polarization experiments tend to cover the frequency range of $\delta \nu/\nu \sim 1$ or even broader in order to monitor the Galactic foreground (e.g. synchrotron and dust) emissions. Achromatic HWP (AHWP) is introduced by Pancharatnam~\cite{pancharatnam} and is employed by EBEX, the balloon-borne CMB experiment~\cite{ebex,matsumura_ahwp}. The stack of the multiple plates with properly chosen offset angles achieves the retardance of $\pi$ broader than $\delta \nu/\nu=0.3$. While this is a very attractive option, one complication with the Pancharatnam AHWP is that the amount of angle rotated by the AHWP becomes an electromagnetic frequency dependence. This effect can be rephrased as the polarization sensitivity axis depends on the instrument bandpass shape and the source spectrum. When the two or more sources are mixed, the uncertainty of the polarization sensitive angle is not only depending on the spectral shapes but also the relative polarized intensities.

In this paper, we introduce the idea to mitigate the spectral dependence of the polarization angle. In section~2, we briefly review the AHWP polarimetery and introduce the mitigation recipe. In section~3, we show the experimental results as a demonstration of the idea. Finally in section~4 we discuss the actual implementations for forthcoming CMB experiments.

\begin{figure}[tb]
\centerline{\includegraphics[width=\columnwidth]{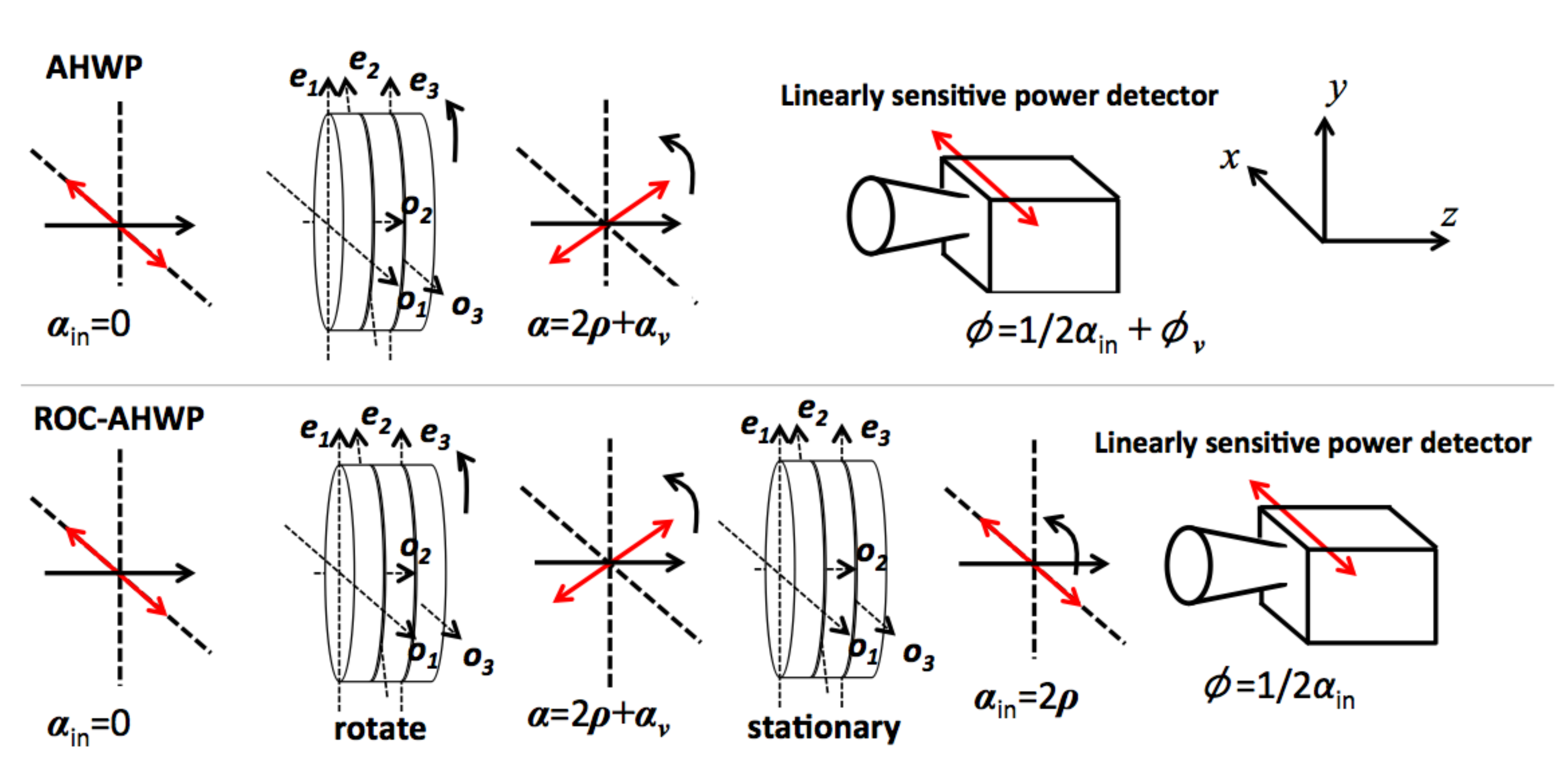}}
\caption{Schematic view of the AHWP polarimeter (top) and ROC-AHWP polarimeter (bottom) using three wave plates ($m=3$).\label{fig:config.pdf}}
\end{figure}

\section{Mitigation recipe}
\subsection{AHWP polarimetry}
The detailed descriptions of the AHWP polarimetry and its formalism can be found in Matsumura et al.~\cite{matsumura_ahwp}. Here we briefly review the AHWP polarimetry.
Figure~\ref{fig:config.pdf} shows the configurations of the polarimeter we assume throughout this paper. 

The expected signal can be formulated by using the Mueller matrix as below 
\begin{eqnarray}
	\vec{S}_{\rm{out}} &=&  G_{x}  \prod_{i=1}^m \left[ R(-\rho-\theta_i) \Gamma(\delta) R(\rho+\theta_i) \right] \ \vec{S}_{\rm{in}},
\end{eqnarray}
where $\vec{S}_{in}$ is the Stokes vector of the incident linearly polarized signal. In this paper, we assume 
\begin{eqnarray}
	\vec{S}_{\rm{in}} &=& \left(I_{\rm{in}},Q_{\rm{in}},U_{\rm{in}},0 \right) \\
	&=& I_{\rm{in}} \left(1, P_{\rm{in}} \cos{2\alpha_{\rm{in}}}, P_{\rm{in}} \sin{2\alpha_{\rm{in}}}, 0\right),
\end{eqnarray}
where $I_{\rm{in}}$ is the incident intensity, $P_{\rm{in}}$ is the incident degree of polarization, $\alpha_{\rm{in}}$ is the incident polarization angle with respect to the detector polarization sensitivity axis, the $x$-axis.
The output Stokes vector is $\vec{S}_{\rm{out}}$. The Mueller matrices, $\Gamma$, $R$, $G_x$ are for a retarder, rotation and wire grid. The detailed matrix elements are shown in Appendix. The angle $\theta_i$ is the offset wave plate angle about the $z$-axis for an $i^{th}$ plate with respect to the $x$-axis, see Figure~\ref{fig:config.pdf}. The total number of wave plates that consist of one set of AHWP is $m$. The angle $\rho$ is the wave plate angle. The retardance $\delta$ is 
\begin{eqnarray}
	\delta = 2\pi\frac{\nu d |n_e-n_o|}{c},
\end{eqnarray}
where $c$ is the speed of light, and $n_o$ and $n_e$ are the ordinary and extra-ordinary indices of the refraction of the wave plate, respectively. As a part of the construction parameters, we determine the thickness of the HWP, $d_c$, as 
\begin{eqnarray}
	d_c = \frac{1}{2} \frac{c}{\nu_c |n_e-n_o|},
\end{eqnarray}
where $\nu_c$ is the central frequency of the detection band.

Without taking into account the effect of reflection from a wave plate, the intensity as a function of a wave plate angle (intensity vs angle = IVA), $I_{\rm{out}}$, for a single HWP can be analytically expressed as 
\begin{eqnarray}
	I _{\rm{out}} &=& \frac{1}{2} \{ I_{\rm{in}} +  \epsilon [Q_{\rm{in}} \cos{4\rho}  +  U_{\rm{in}} \sin{4\rho} ] \} \nonumber  \\
	             &=& \frac{1}{2} [ I_{\rm{in}} +  \epsilon \sqrt{Q_{\rm{in}}^2+ U_{\rm{in}}^2} \cos{(4\rho - 2\alpha_{\rm{in}} )} ], 
	             	\label{eq:model_iva}
\end{eqnarray}
where $\epsilon$ is modulation efficiency defined as 
\begin{eqnarray}
	\epsilon \equiv \frac{ I_{p_{\rm{out}}} }{  I_{p_{\rm{in}}}}, \ \ \mbox{where} \ I_{p} = \sqrt{Q^2 + U^2}.
	\label{eq:modeff}
\end{eqnarray}
We also define the phase of IVA and relate to the incident polarization angle as 
\begin{eqnarray}
	\phi = \frac{1}{2}\alpha_{\rm{in}}.
	\label{eq:phase}
\end{eqnarray}
When we apply Equation~(\ref{eq:model_iva}) to the IVA of AHWP, we introduce an extra phase, $\phi_\nu$, as 
\begin{eqnarray}
	I _{\rm{out}} &=& \frac{1}{2} [ I_{\rm{in}} +  \epsilon \sqrt{Q_{\rm{in}}^2+ U_{\rm{in}}^2} \cos{(4\rho - 2\alpha_{\rm{in}} - 4\phi_{\nu} )} ]
	\label{eq:fit_iva}
\end{eqnarray}
in order to take into account the phase variation as a function of frequency, and thus 
the phase of the modulation is expressed as 
\begin{eqnarray}
\phi = \frac{1}{2} \alpha_{\rm{in}} + \phi_{\nu}.
\end{eqnarray}
The extra term, $\phi_\nu$, has a spectral dependence, i.e. the spectra of a source and instrument. Thus, a polarimeter using AHWP has the intrinsic source of uncertainty in its polarization angle unless one knows the source spectrum to the required precision. The quantitative discussion about this effect in the context of a CMB polarization experiment can be found in Matsumura et al. and Bao et al.\cite{matsumura_ahwp,bao}.

\subsection{Introducing the rotational-offset-canceling AHWP}
The idea to mitigate the frequency dependence of the polarization angle with the use of rotating AHWP is to place a second set of the AHWP that is moving with respect to the first set of the AHWP. The simplest configuration for the second set is to prepare an identical stationary AHWP as shown in the bottom of Figure~\ref{fig:config.pdf}. 

The first set is placed to modulate the incident polarization angle. The second set still maintains the high modulation efficiency and also rotate the offset of the incident polarization angle back. Hereafter, we call this configuration the rotational-offset-canceling AHWP (ROC-AHWP). The above configuration can be written down by using the Mueller matrices as
\begin{eqnarray}
	\vec{S}_{\rm{out}} &=&  G_{x} \prod_{i=1}^m \left[ R(-\rho_1-\theta_i) \Gamma(\delta) R(\rho_1+\theta_i) \right] \times  \nonumber \\ && \prod_{i=1}^m \left[ R(-\rho_2-\theta_i) \Gamma(\delta) R(\rho_2+\theta_i) \right] \ \vec{S}_{\rm{in}},
	\label{eq:roc_ahwp_mueller}
\end{eqnarray}
The detected intensity is the first element of the output Stokes vector. The rotational angles of the two AHWPs, $\rho_1$ and $\rho_2$, need to be not identical. The simplest configuration is to set $\rho_1=0$ and let $\rho_2$ to rotate. We reconstruct the phase, $\phi_\nu$, from IVA and this now does not have the spectral dependence while maintaining high modulation efficiency, $\epsilon$.

\begin{figure}[tb]
\centerline{\includegraphics[width=\columnwidth]{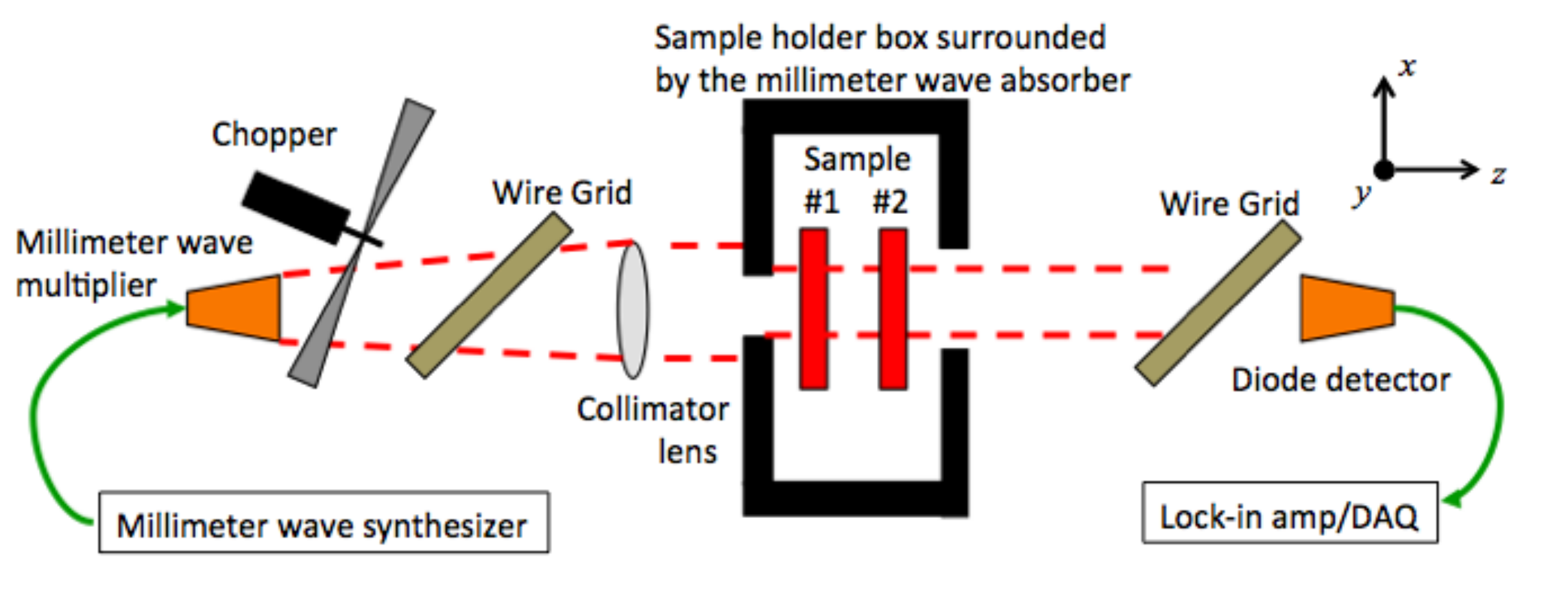}}
\caption{Schematic view of the measurement setup. \label{fig:setup.pdf}}
\end{figure}

\section{Measurements}
We conducted an experiment to demonstrate the concept of the ROC-AHWP. 

\subsection{Sample preparation}
We prepare the 50~mm diameter of an A-cut single crystal sapphire as a wave plate. We construct the two sets of the stacked three wave plates ($m=3$) with the offset angles of (0, 55, 0) degrees. The relative offset angles between the wave plates are aligned within 0.5 degrees. The thickness of each sapphire sample is $3.82\pm0.005$~mm. The ordinary and extra-ordinary indices of the sapphire at the room temperature are assumed to be 3.07 and 3.40~\cite{brad_thesis}. The surfaces of the first and the third sapphire layers are anti-reflection (AR) coated with two layers of dielectric materials. The first AR layer on the sapphire is Stycast2850FT with a thickness of $0.24$~mm and the second layer on Stycast2850FT is the teflon sheet with a thickness of $0.4$~mm. We apply this two-layer AR coating only to the sapphire surface that faces to the air and did not apply the interface layer between the two sapphires. The three layers of sapphire are held mechanically.   

\subsection{Experimental setup}
Figure~\ref{fig:setup.pdf} shows the experimental setup to measure the modulation efficiency and the phase of IVA. 
We use the millimeter-wave generator and the $\times 6$ (W-band) and $\times 9$ (D-band) multipliers as millimeter wave sources to produce the spectral range between 72 and 162~GHz. The source is linearly polarized and we also use two free-standing wire-grid polarizers to define the polarized orientation. We define the incident polarization angle of $\alpha_{in} = 0$~degrees as the polarization vector referenced with respect to the $x$-axis. The linearly polarized signal is measured by a diode detector that is sensitive to linearly polarized light. The signal is chopped at 13~Hz and the detected modulated signal is demodulated by the lock-in amplifier and read by data acquisition system.

The incident beam from the source is collimated by a spherical lens made by Rexolite and the beam is further collimated by the 2.5~cm diameter aperture. The AHWP (sample \#1) is located on the automated rotatable mount and the stationary AHWP (sample \#2) is placed after the rotating AHWP. 
The first AHWP rotated over 360 degrees with a step of 10 degrees at the given electromagnetic frequency and we extract the modulation efficiency and phase from the IVA. We repeat this measurement over the frequency range from 72 to 108 GHz using the W-band source and detector, and from 117 to 162 GHz using the D-band source and detector. 

The AHWP is mounted on a sample holder. The rotational position of the sample holder is aligned with respect to the $x$-axis within 5 degrees. We calibrate the global offset angle about the $z$-axis between the sample and the $x$-axis by using the wire grid.

\begin{figure}[tb]
\centerline{\includegraphics[width=\columnwidth]{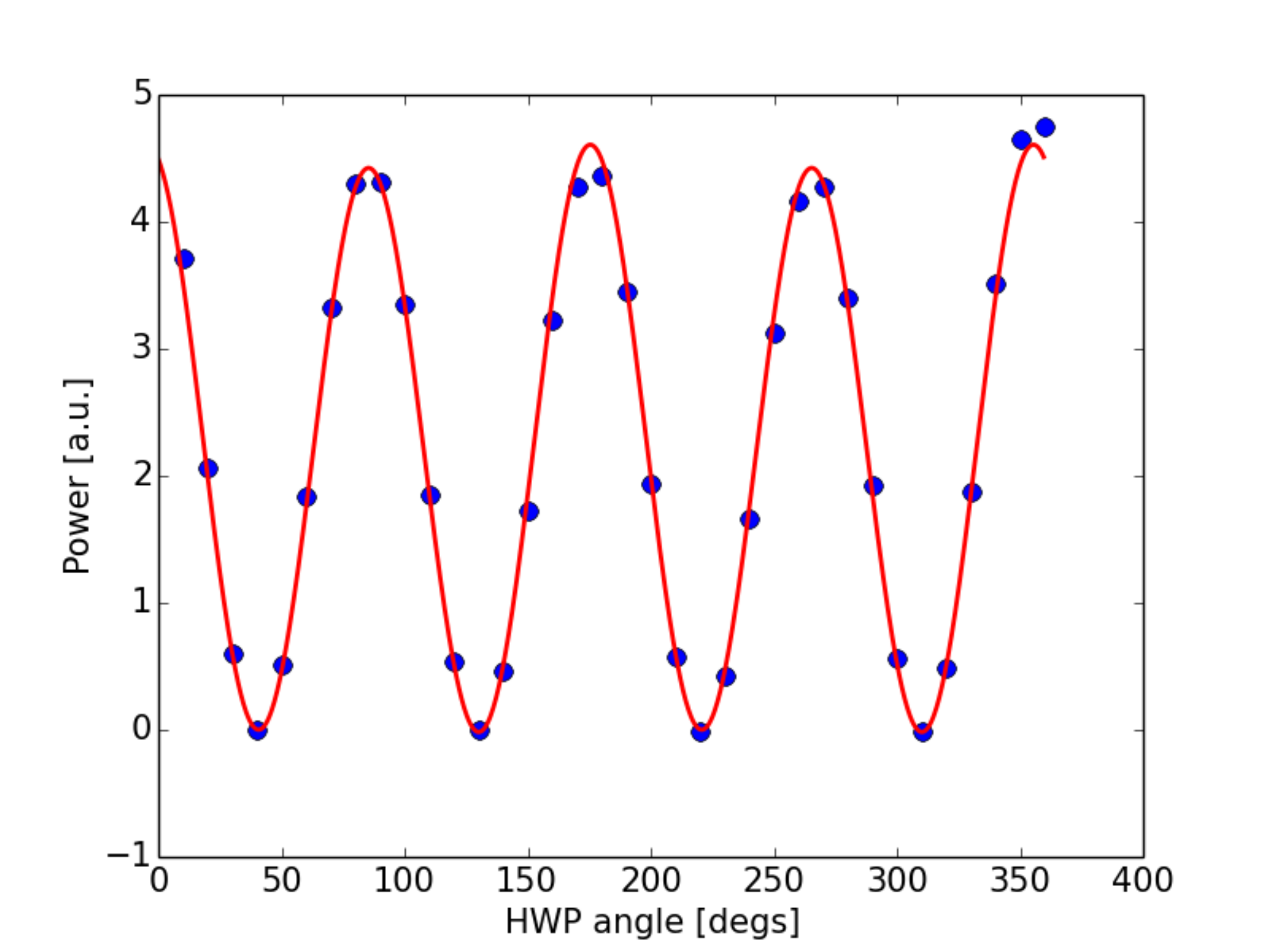}}
\caption{Typical IVA data and the fit using Equation~(\ref{eq:fit_iva}).\label{fig:IVA_example.pdf}}
\end{figure}

\begin{figure}[tb]
\centerline{\includegraphics[width=\columnwidth]{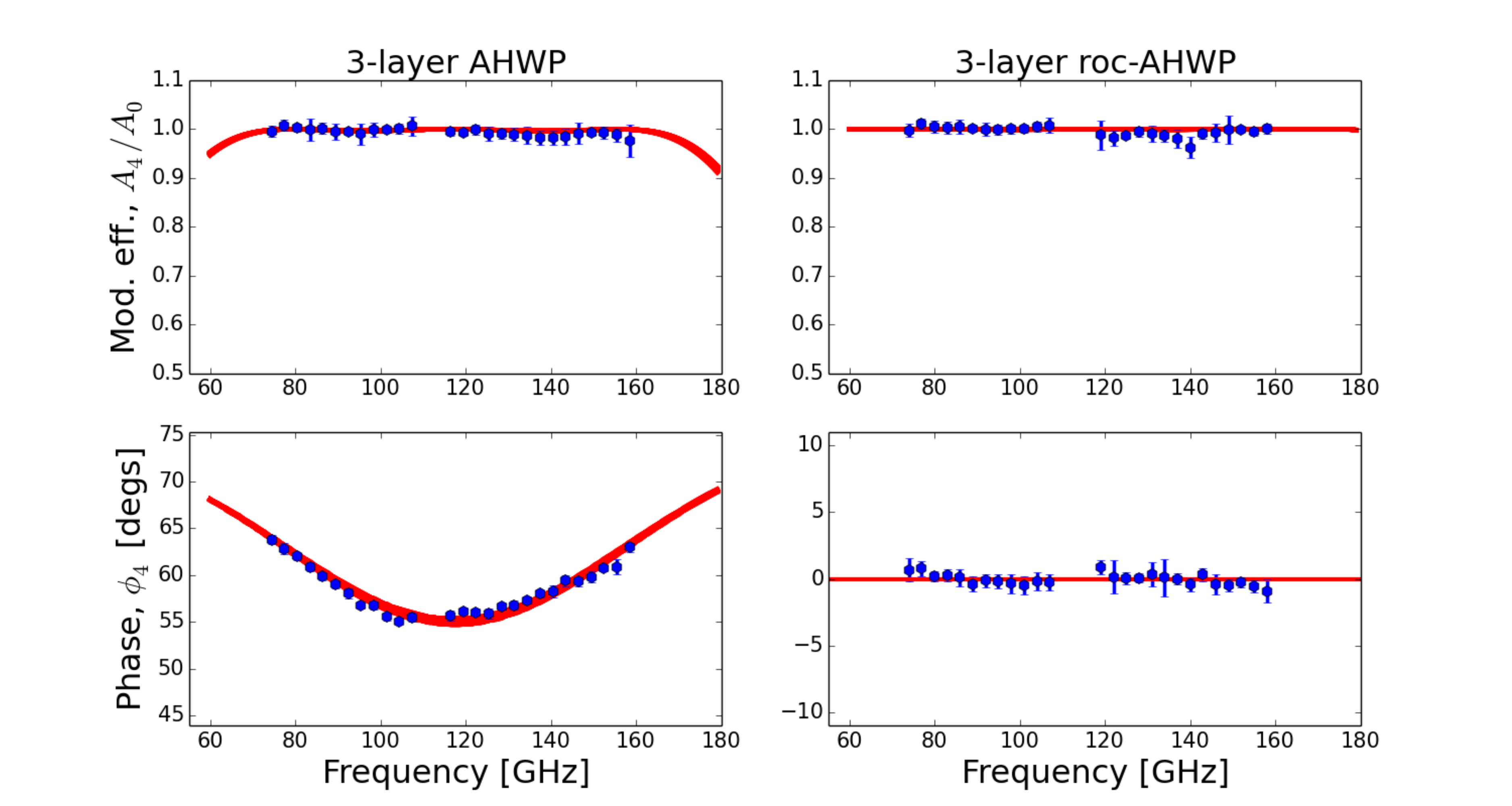}}
\caption{The modulation efficiency (top) and the phase (bottom) as a function of the incident electromagnetic frequency for the three-layer AHWP (left) and ROC-AHWP (right). The blue points are measured data and the red lines are not a fit but predicted curves. The incident polarization angle is $\alpha_{in}=0$ degrees. \label{fig:PoleffPhase_4panels}}
\end{figure}

\section{Results}
Figure~\ref{fig:IVA_example.pdf} shows the typical IVA data. The IVA is fitted with the following model,
\begin{eqnarray}
	I_m  &=& A_0 + A_{2} \cos{(2\rho + 2\phi_2)} + A_{4} \cos{(4\rho + 4\phi_4)}. 
	\label{eq:fit_iva}
\end{eqnarray}
We define the normalized modulation efficiency, $\tilde{\epsilon}$, and the phase, $\phi$,  of the IVA as 
\begin{eqnarray}
	\tilde{\epsilon} \equiv \frac{A_4}{A_0}, \ \ \ \ \phi \equiv \phi_4
\end{eqnarray}
The normalized efficiency for a single HWP can be analytically solved as 
\begin{eqnarray}
	\tilde{\epsilon} = \frac{P_{out}}{P_{in}}.
\end{eqnarray}
We use the normalized modulation efficiency instead of the modulation deficiency defined in Equation~(\ref{eq:modeff}). This is because the source intensity, $I_{p_{in}}$, is not well known over the spectral range while we can ensure to achieve $P_{in}=1$ by using the combination of the linearly polarized source and the wire grid.


Figure~\ref{fig:PoleffPhase_4panels} shows the modulation efficiency and the phase as a function of the electromagnetic frequency for the AHWP and the ROC-AHWP. 
The phase of the AHWP shows the spectral dependence. On the other hand, the phase of the ROC-AHWP has a flat response. The modulation efficiency of AHWP and ROC-AHWP is close to 1 over broadband, $\delta \nu/\nu \sim 0.7$.

The error on each data point in Figure~\ref{fig:IVA_example.pdf} and \ref{fig:PoleffPhase_4panels} is estimated by three noise sources, statistical noise, thermal drift, and standing wave in the system due to the non-perfect AR coating. The fractional statistical error is less than $1\times10^{-3}$ over 1~sec of the integration time and the temperature of the source is stationary within 0.1~degrees. The thermal drift is estimated by repeating the measurements multiple times over a few hour separations. The systematic error due to the reflection is estimated by the deviation of IVA from Equation~(\ref{eq:fit_iva}). 

The red lines are the predicted curves without taken into account the effect of reflection from the wave plate surfaces. The predicted curves have thickness due to the uncertainties in the thickness of each wave plate and the offset angles between the wave plates. The data points and the predicted curves are well agreed within 1-$\sigma$ error bar. 

\begin{figure}[tb]
\centerline{\includegraphics[width=\columnwidth]{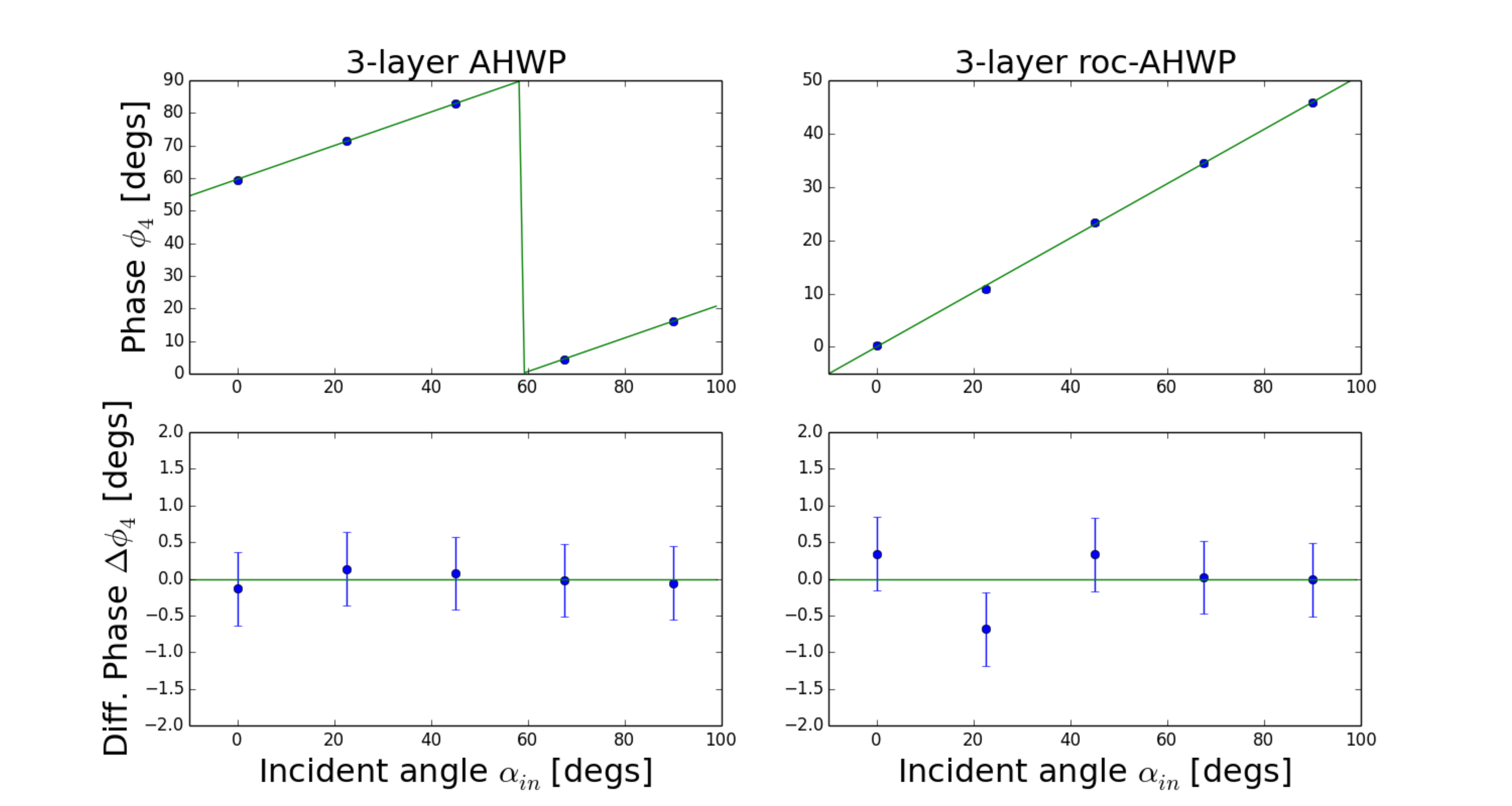}}
\caption{The band averaged phase as a function of the incident polarization angle. The solid line is a linear fit that is wrapped around between 0 and 90 degrees.\label{fig:Phase_Alpha_summary}}
\end{figure}

Figure~\ref{fig:Phase_Alpha_summary} shows the phase as a function of the incident polarization angle. The phase at each incident angle is averaged between 72 and 108 GHz. 
The error bar on each data point is dominated by the uncertainty of the incident polarization angle. The uncertainty of the wire grid orientation is $\pm0.5$~degrees and this is limited by the tick mark of the wire grid holder. This error is propagated to the error on the phase.
\begin{table*}[tb]
   \centering
   \begin{tabular}{c||c|c|c|c|c} 
	Experiment & \# of plates & $d_c$  [mm] ($\nu_c$ [GHz]) & Band [GHz] & Bandwidth [\%] & HWP offset angles [degs] \\
\hline\hline
	POLARBEAR-2 & 3 & 3.74 (126.6) & 95, 150 & 30 & (0, 56, 0) \\ 
\hline
	LiteBIRD & 9 & 2.52 (187.5)  & 60, 78, 100   & 23 & (0, 18.5, 217.5, 73.9, 141.5, \\ \cline{4-5}
	                  &     &             & 140, 195, 280 & 30 &  \ \ \ \ 73.9, 217.5, 18.5, 202.7)  \\ 
   \end{tabular}
   \caption{The design parameters for ROC-AHWP for POLARBEAR-2 and LiteBIRD. The thickness is for a single plate and the variable, $\nu_c$, corresponds to the frequency that has a retardant of $\pi$.}
   \label{tab:hwp_pars}
\end{table*}

\section{Discussions}
We discuss the detailed feature that may become the potential systematic effect with the use of ROC-AHWP. We also discuss the design parameters for two 
upcoming CMB experiments that potentially use the AHWP base polarimetry.

\subsection{Incident angle dependent modulation efficiency}
When one designs the AHWP, there is a tradeoff between the high modulation efficiency and the bandwidth by tuning the offset angles. In one application, if one does not necessary need the modulation efficiency of 1 in the detection bandwidth, one can gain wider bandwidth with modulation efficiency less than 1. 

As for the ROC-AHWP design, one needs a careful treatment in this tradeoff. When the retardance is not equal to $\pi$ after the first set of the AHWP, a partial circularly polarized light is incident to the second AHWP. In such a case, the outgoing light is further circularly polarized. Therefore, the second AHWP enhances to the lower modulation efficiency.

If the retardance is $\delta\neq\pi$, the polarized light after the first AHWP gains some amount of $V \neq 0$. This non-zero $V$ polarized light incidents to the second AHWP with some polarization angle with respect to the axis of the second AHWP. In such a case, the degradation of the modulation efficiency depends on what the initial polarization angle is with respect to the first AHWP. If one designs the ROC-AHWP for an application that needs no incident angle dependence in the modulation efficiency, one has to carefully choose the construction parameters in order to maximize the modulation efficiency for all the incident polarization angles with some sacrifice of the bandwidth.

\subsection{ROC-AHWP design for POLARBEAR-2 and LiteBIRD}
We describe the candidate ROC-AHWP design for upcoming CMB polarization experiments, POLARBEAR-2 and LiteBIRD. Our design assumes the design parameters described in Table~\ref{tab:hwp_pars}. We optimize the offset angles in such that the modulation efficiency is as close as $\sim1$ and the flatness of the phase over the detection bands.

\begin{figure}[tbph]
\centerline{\includegraphics[width=\columnwidth]{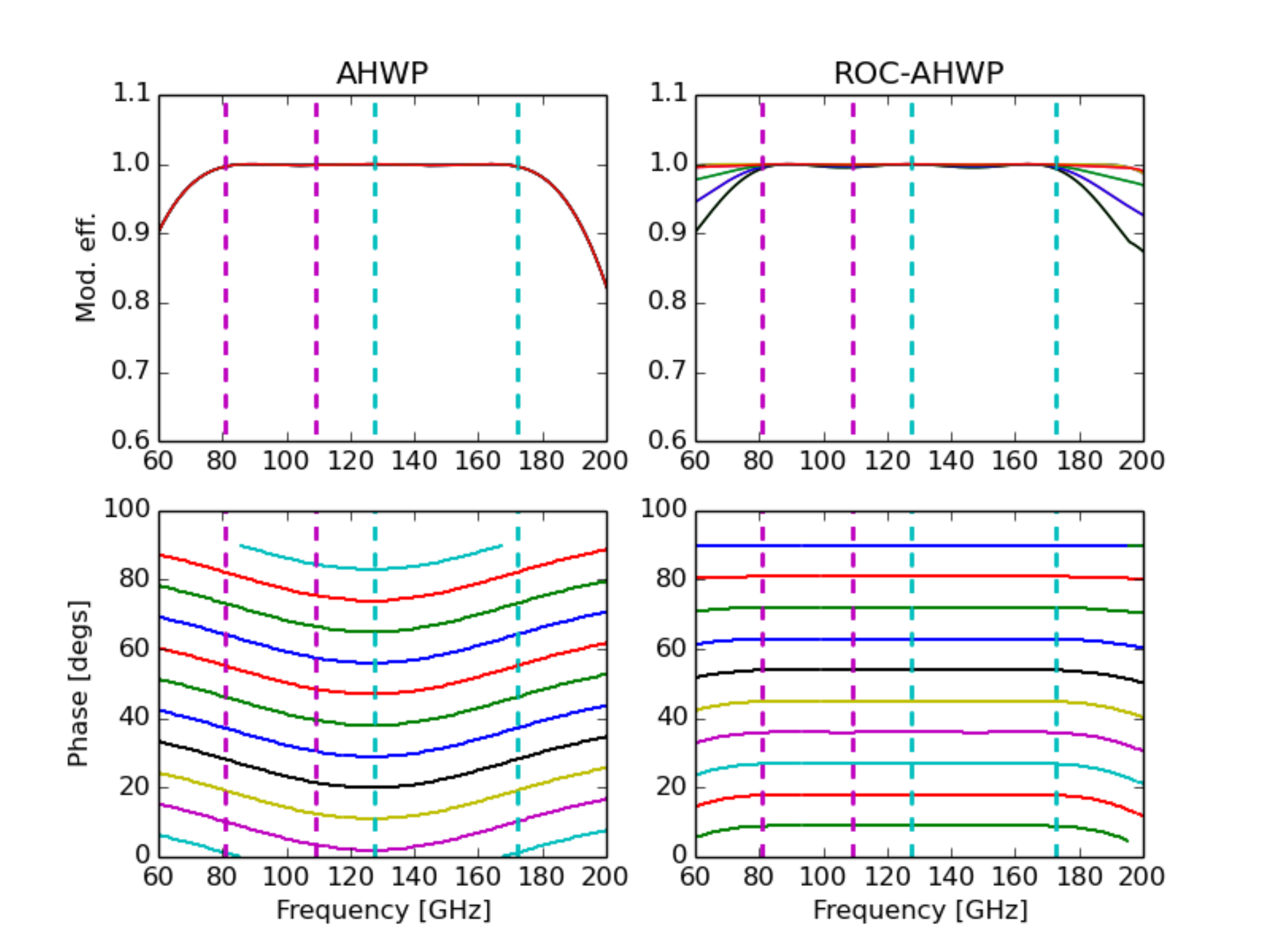}}
\caption{The simulation results of the modulation efficiency and the phase for the three layer AHWP and ROC-AHWP. The parameters are tuned for POLARBEAR-2 specification. The dashed vertical lines indicate the 30~\% bandwidth of 95~GHz and 150 GHz. The different color corresponds to the different incident angle, $\alpha_{\rm{in}}$, in 18 degree step. \label{fig:PB2_Ip.pdf}}
\end{figure}

Figure~\ref{fig:PB2_Ip.pdf} shows that the modulation efficiency and the phase of the AHWP and ROC-AHWP for POLARBEAR-2. The band averaged modulation efficiencies for 95 and 150 GHz (with 30\% bandwidth) of the ROC-AHWP design are $>99.8~\%$ and $ > 99.4~\%$, respectively, for all the incident polarization angles. The RMS variations of the phase within the detection bandwidth of 95 and 150~GHz are less than 0.05 degrees. 

The effect, which is previously discussed as the incident angle dependent modulation efficiency, is particularly evident below 80~GHz and above 173~GHz of the ROC-AHWP. Some amount of this effect is also present within the detection band indicated as dashed lines.  
The impact of these values to a CMB polarization experiment is beyond the scope of this paper, and it requires the end-to-end simulation to assess the impact of this performance. When the requirement is stringent, the potential solution to improve the performance is to add more plates to widen the bandwidth with higher modulation efficiency.

\begin{figure}[tbph]
\centerline{\includegraphics[width=\columnwidth]{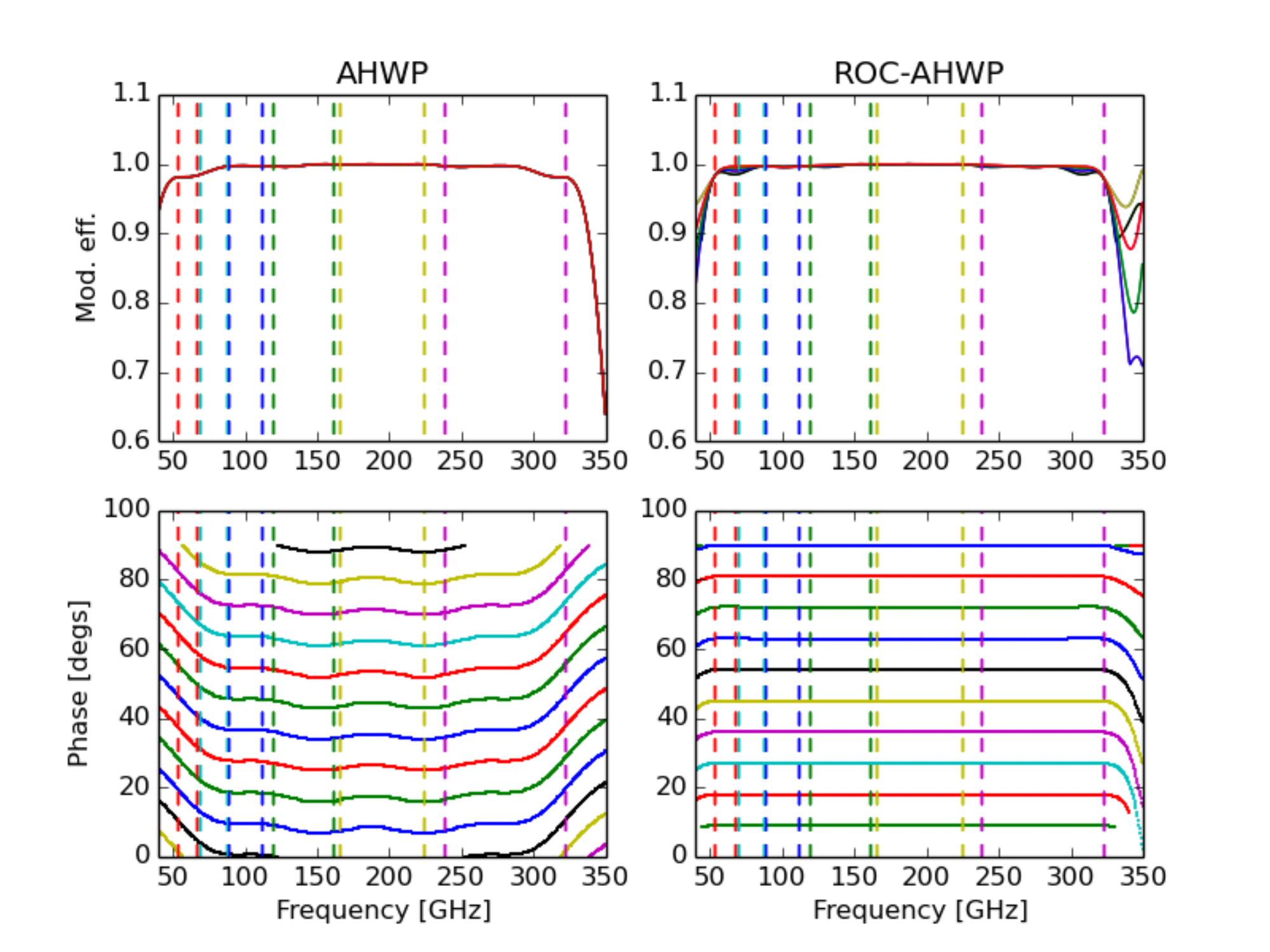}}
\caption{The simulation results of the modulation efficiency and the phase for the nine layer AHWP and ROC-AHWP. The parameters are tuned for LiteBIRD specification. The dashed vertical lines indicate the 23~\% bandwidth for 60, 78, and 100~GHz bands and 30~\% bandwidth of 140, 195 and 280~GHz. The different color corresponds to the different incident angle, $\alpha_{\rm{in}}$, in 18 degree step. \label{fig:LB_Ip.pdf}}
\end{figure}

Figure~\ref{fig:LB_Ip.pdf} shows the modulation efficiency and the phase of the AHWP and ROC-AHWP for LiteBIRD. It consists of the 9 stacked sapphire plates. 
The offset angles for the 9 plates are chosen by conducting the Monte Carlo simulation and maximizing the modulation efficiency over $50-320$~GHz. This is one of the candidate offset angles that cover the detection bandwidth of LiteBIRD. 
Table~\ref{tab:lb_modeff_phase} shows the band averaged modulation efficiency and the RMS fluctuation of the phase within the band.

\begin{table}[tb]
   \centering
   \begin{tabular}{c||c|c} 
	Band [GHz] & $\epsilon$ & $\phi_{rms}$ [degs]  \\
\hline\hline
	60 & $0.991^{+0.002}_{-0.004}$  & $1.2\times10^{-3}$\\
\hline
	78 & $0.995^{+0.002}_{-0.003}$ &  $1.6\times10^{-3}$  \\
\hline
	100 & $0.996^{+0.001}_{-0.000}$ &  $0.0$  \\
\hline
	140 & $0.999^{+0.001}_{-0.001}$ &  $5\times10^{-4}$  \\
\hline
	195 & $1.000^{+0.000}_{-0.000}$ &  $6\times10^{-4}$  \\
\hline
	280 & $0.995^{+0.001}_{-0.002}$ & $1.8\times10^{-3}$ \\
   \end{tabular}
   \caption{The design parameters for ROC-AHWP for POLARBEAR-2 and LiteBIRD. The thickness is for a single plate and the variable, $\nu_c$, corresponds frequency that has a retardant of $\pi$.  The quoted error in modulation efficiency is from the incident angle dependence.  \label{tab:lb_modeff_phase}}
\end{table}

\subsection{Implementations}
This ROC-AHWP inevitably doubles the total thickness of the AHWP in the optical path and the surface that potentially reflects the incident radiation. In the application that needs high transmittance, one has to choose a low-loss birefringent material and proper anti-reflection coating in the required bandwidth. 

When the two parallel optical elements are placed like the ROC-AHWP, the standing wave may cause undesired fringe pattern in the spectrum of the detection band. The simplest solution is to apply a proper AR coating. As a result, the requirement to the broadband AR coating may become more demanding with the ROC-AHWP.

The modulation efficiency and phase response at around $\epsilon\sim1$ and $\Delta\phi\sim0$ behave the same regardless of the order of the first AHWP being rotating and the second set being stationary, or vice versa. One can choose the configuration that fits better to the optical system. This feature can be considered as redundancy in a system requiring a polarimeter on balloon-borne and satellite missions. Generally the rotation of the polarimeter is one of the key features to maintain the measurements and the failure of the rotation can be fatal.
The second stationary AHWP can hold the rotational mechanism that is nominally not activated and the extra-mechanism can be activated when there is a failure in the rotation of the first AHWP.

In millimeter wave applications, sapphire is one of the typical birefringent materials due to the large differential indices between the ordinary and extra-ordinary axes. Sapphire is known as a high thermal conductive material ($>$1~W cm$^{-1}$K$^{-1}$ between 4 and 100~K)~\cite{pobell}. Therefore, the stationary sapphire with a proper AR coating can serve not only as a part of the second set of the AHWP in the ROC-AHWP but also serve as a thermal filter that absorbs infrared radiation at cryogenic temperature. For this use, the correct choice of the AR coating material with features, absorptive at IR and transparent in millimeter wave (such as a thin layer of Stycast), should achieve a functionality as a thermal filter. The similar concept is proposed using alumina with the Stycast AR coatings as an IR filter at the cryogenic temperature~\cite{alumina_filter}.

In a HWP based polarimetry using a continuous rotation, another possible operational mode is to rotate the two AHWPs of the ROC-AHWP in opposite directions. In Equation~(\ref{eq:roc_ahwp_mueller}), one can set $\rho_1 = \omega t$ and $\rho_2 = - \omega t$, where $\omega$ is the angular velocity of the AHWP and $t$ is time. In this mode, the total angular momentum of the ROC-AHWP can be canceled. When the angular momentum due to the rotation of the AHWP can be a source of disturbance in a system, such as a satellite or balloon payload attitude control system, one can achieve the continuously rotating HWP based polarimetry with zero-momentum. In this case, one should aware that the polarization signal  appears at $8\omega t$ instead of $4\omega t$, i.e. the nominal frequency that the polarization signal appears with a single AHWP. Therefore, this double spinning ROC-AHWP in the opposite directions can add an extra feature of increasing the signal band. This is a particularly interesting option when a physical rotational frequency of the AHWP is limited.

While we demonstrated this concept of the ROC-AHWP at millimeter wavelengths experimentally, the principle of the ROC-AHWP can be applicable to any wavelengths. Thus, this ROC-AHWP can be a viable candidate technology for any applications that use broadband polarimetry, including solar physics, expoplanet search, aerosol characterization, and biomedical applications. 
   
\section{Conclusions}
We introduced the recipe to mitigate the spectral dependent phase response of the AHWP polarimeter by introducing the second set of the AHWP that rotates or is stationary with respect to the first set of the AHWP. 
We experimentally show that the ROC-AHWP configuration with $m=3$ achieves $\Delta \phi_\nu < 1$~degrees while the phase of a single set of AHWP varies $\Delta \phi_\nu \sim 10$~degrees between 72 and 162~GHz. We also computationally show the potential ROC-AHWP design for POLARBEAR2 with $m=3$ and for LiteBIRD with $m=9$. 
Although we demonstrate this ROC-AHWP at the millimeter wavelength, the ROC-AHWP concept is applicable to any broadband polarimeter application.

\section*{Acknowledgement}
We would like to thank to Masashi Hazumi for useful discussions. This work was supported by MEXT KAKENHI (Grant Nos. 24740182 and 24111715).

\section*{Appendix}
\subsection*{Elements of Mueller matrix}
The elements of the Mueller matrix that are used in this paper are listed below.
\begin{eqnarray}
\Gamma(\delta) &=& 
\left[ \begin{matrix}
  1 & 0 & 0 & 0 \\
  0 & 1 & 0 & 0 \\
  0 & 0 & \cos{\delta} & -\sin{\delta} \\
  0 & 0 & \sin{\delta} & \cos{\delta}
 \end{matrix} \right]  \\
R(\theta) &=&
\left[ \begin{matrix}
  1 & 0 & 0 & 0 \\
  0 & \cos{\theta} & -\sin{\theta} & 0 \\
  0 & \sin{\theta} & \cos{\theta} & 0 \\
  0 & 0 & 0 & 1
 \end{matrix} \right] \\   
G_x &=& \frac{1}{2} 
\left[ \begin{matrix}
  1 & 1 & 0 & 0 \\
  1 & 1 & 0 & 0 \\
  0 & 0 & 0 & 0 \\
  0 & 0 & 0 & 0
 \end{matrix} \right].
\end{eqnarray}
For more general expression, please see Appendix in Shurcliff~\cite{shurcliff}.

\end{document}